%% file: main.tex
\documentclass[sigplan,screen,pbalance]{acmart}
\AtBeginDocument{%
  }

\usepackage{algorithm2e}
\usepackage{enumitem}
\usepackage{listings}
\usepackage{tcolorbox}
\usepackage[dvipsnames]{xcolor}
\usepackage{balance}
\usepackage{multirow}
\usepackage{multicol}
\usepackage{pifont}
\usepackage{xcolor}

\newcommand{\Vy}{{\color{green!60!black}\ding{51}}} 
\newcommand{\Xn}{{\color{red!80!black}\ding{55}}}   

\def\Tech/{\textsc{ReDig}}

\begin{document}

\title{Refining LLM-based Directed Test Input Generation via Runtime Value Feedback}
\author{Narin Han}
\email{narinhan@cbnu.ac.kr}
\affiliation{%
  \institution{Chungbuk National University} 
  \city{Cheongju}
  \country{Republic of Korea}  
}

\author{Shin Hong}
\email{hongshin@cbnu.ac.kr}
\affiliation{%
  \institution{Chungbuk National University} 
  \city{Cheongju}
  \country{Republic of Korea}  
}


\begin{abstract}
Despite their remarkable capabilities, LLM-based directed 
input generation produces off-target test inputs when their underlying hypotheses 
on the target program's behaviors are incorrect.
To effectively ground LLM-based directed input generation in 
actual program behaviors, we propose \Tech/, a runtime feedback-guided
framework for iterative refinement of LLM-based directed input generation.
When a generated input fails to reach a given target code location, \Tech/ conducts
a postmodern analysis to identify incorrect hypotheses by comparing 
expected runtime values and their actual runtime values observed in 
the off-target test execution.
\Tech/ utilizes the postmortem analysis result as feedback to refine 
the inference context, guiding the subsequent test generations toward the target location.
In this paper, we present our work-in-progress on \Tech/. We illustrate 
how runtime feedback-guided refinement operates through four case studies 
with two real-world C/C++ programs ({\tt Poppler} and {\tt Libsndfile}).
Furthermore, we report preliminary experiment results that show
\Tech/ successfully generates crash-reproducing inputs 
for 24 out of the 59 CVEs collected from \textsc{Sec-Bench}.
%



\end{abstract}

\keywords{Test Input Generation, LLM-based Directed Input Generation, Runtime Monitoring}




\setcopyright{none}
\settopmatter{printacmref=false}
\renewcommand\footnotetextcopyrightpermission[1]{}
\pagestyle{plain}
\maketitle

\input{introduction}
\input{technique}
\input{casestudy}

\input{experiment-secbench}

\input{discussion}

\section{Conclusion}
\label{sec:conclusion}
We preset our work-in-progress on \Tech/, a runtime value feedback-guided refinement framework for 
LLM-based directed input generation. 
\Tech/ uses runtime values observed from off-target executions to 
refine the information provided to the LLM directed input generator. 
The preliminary experiments with 59 CVEs from real-world C/C++ programs show
that \Tech/ effectively generates directed test input generation utilizing
the feedback information.


\bibliographystyle{ACM-Reference-Format}
\bibliography{references}

\end{document}
\endinput

%% file: introduction.tex
\section{Introduction}
LLM-based directed input generation~\cite{jiang2024towards, vikram2026fuzzing} 
synthesizes test scripts that produce inputs intended to reach specific program 
locations designated as targets\footnote{
Hereafter we refer to a test input produced by a LLM generated script 
as a ``generated test input'' for simplicity. 
}. 
This approach leverages the abilities of LLMs to infer the input constraints required 
to reach target locations based on the available code context and pretrained 
knowledge of relevant open-source projects and input file formats. 
It has shown particular promise for generating proof-of-vulnerability inputs 
for confirming the reachability of vulnerable code identified by 
static analyses~\cite{codamosa} or LLM-based code analyses~\cite{atlantis}.

However, the effectiveness of LLM-based directed input generation is 
constrained when the available code does not provide sufficient 
contextual information~\cite{testeval}. 
Even with adequate context, LLMs may still fail to
produce effective inputs due to plausible-but-incorrect 
assumptions about program behavior~\cite{codeaugur}. 
Therefore, 
reliable directed input generation requires mechanisms to 
ground the process in observed runtime behavior rather than relying solely on LLM inference.

We propose \Tech/, a runtime feedback-guided refinement 
framework for LLM-based directed input generation,
and present the preliminary results of our work-in-progress.
\Tech/ adds a control loop around an LLM-based 
directed input generation technique to 
refine the information
given to the LLM input generator using observations from prior test executions.
To this end, when a generated input fails to reach the target location, 
this framework provides the LLM test generator with runtime observations 
extracted from the failed execution via an interactive debugger (e.g., {\tt gdb}~\cite{gnugdb} and {\tt lldb}~\cite{LLVM:CGO04}), enabling 
it to revise incorrect assumptions about program behaviors,
and generate a more effective directed test input in the next iteration. 

We first explore the feasibility of the proposed technique and elaborate how it works through 
four case studies drawn from {\tt Libsndfile} and {\tt Poppler} (see Section~\ref{sec:casestudy}). 
These case studies describe how LLMs can effectively use runtime value 
feedback to iteratively explore different paths, and 
reach uncovered target locations. These results  
suggest that LLMs can use runtime value feedback effectively
to diagnose why the previous test script failed to reach the target, and 
leverage this knowledge in subsequent directed input generation.

We further evaluate \Tech/ on generating failing inputs that reproduce 
given crash information (see Section~\ref{sec:experiment}). 
With 59 CVEs from \textsc{Sec-Bench}~\cite{sec-bench}, 
we assessed whether \Tech/ can effectively generate crash-reproducing inputs. The results show that \Tech/ generated crash-reproducing inputs successfully for 24 CVEs.

%% file: technique.tex
\section{Technique}
\subsection{Overview}

\Tech/ extends LLM-based directed input generation to 
an feedback-guided iterative refinement process.
We posit that, based on its implicit knowledge on 
the language semantics and target program functionalities, 
a LLM internally hypothesizes a set of program state conditions 
under which the target line can be covered, and
a set of input conditions required for reaching the expected 
program states.
Accordingly, an off-target test input is resulted by
wrong hypotheses derived in the inference.
Just as a faulty code at unexpected location leads
an program execution to reach an unintended program state, 
a wrong hypothesis derived in a middle leads the inference result to 
an unintended, off-target test input.

From this perspective, we conjecture that comparing hypothesized program states 
with actual runtime states from off-target execution 
can reveal erroneous hypotheses and guide context refinement.
Furthermore, inspired by the LLM-driven scientific debugging~\cite{AutoSD}, 
we conjecture that a LLM can effectively suspect suspicious 
hypotheses, proposes runtime value tests to validate whether 
the suspected hypotheses are actually counterfactual, and suggest
alternative hypotheses incorporating the facts from the observed 
runtime values.
We implement this idea as {\it runtime-value feedback generation}
and {\it feedback-guided iterative refinement} in \Tech/.
Once an off-target input is generated, \Tech/ conducts postmortem
incorporated with an interactive debugger (i.e., {\tt gdb}).
We expect that iterating test input generation with refinement 
effectively grounds the underlying hypotheses in the actual target
program behaviors, and guides the test generation process to 
generating on-target inputs.

\begin{figure}[t!]
    \centering
    \includegraphics[width=\columnwidth]{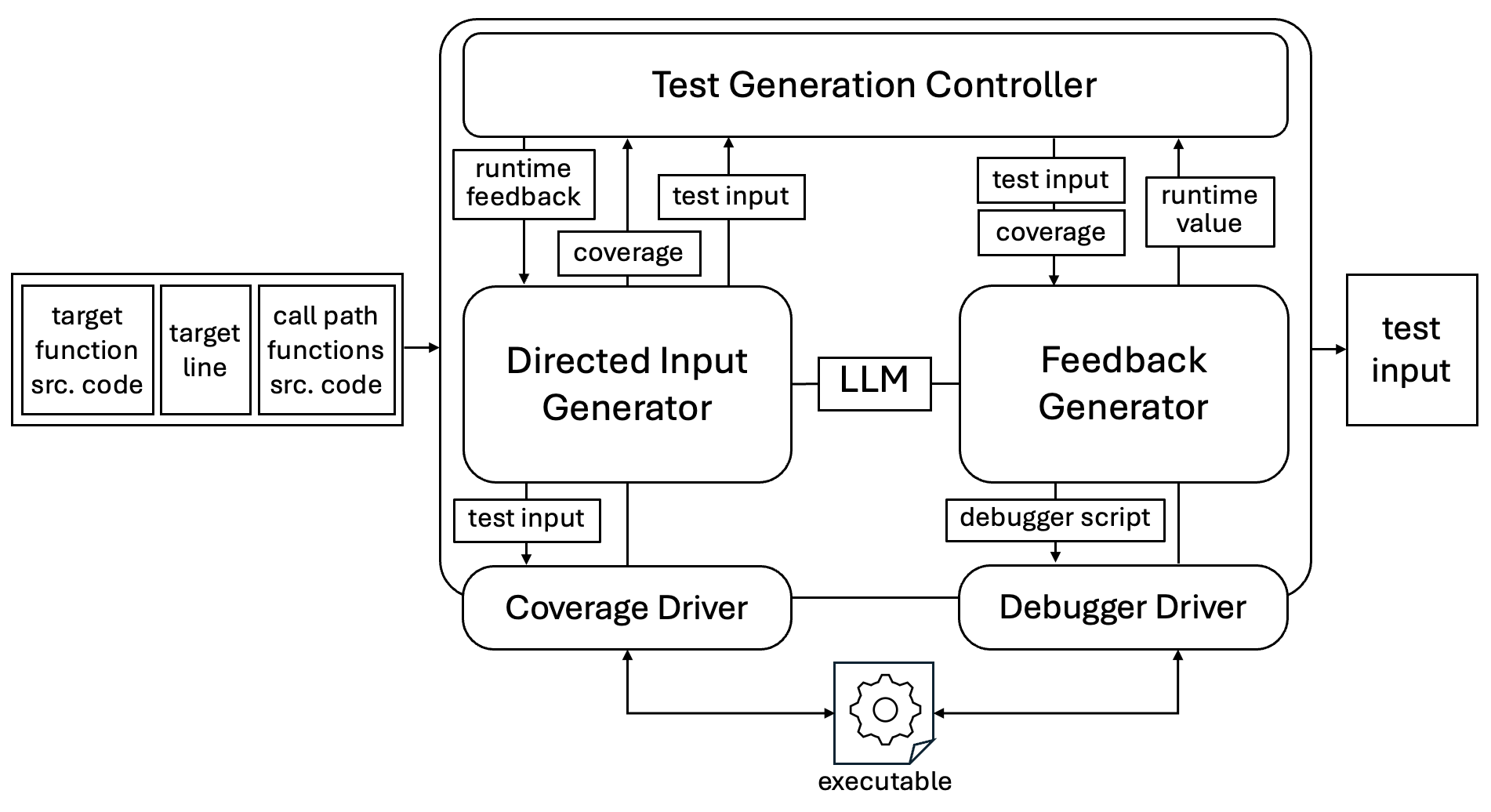}
    \caption{\Tech/ Framework}
    \label{fig:framework}
\end{figure}

Figure~\ref{fig:framework} describes three major components
and their interactions. 
The {\it Directed Input Generator} drives the LLM to generate
directed test inputs, and evaluate whether their executions
reach the target line (see Section~\ref{sec:dig}) via {\it 
Coverage Driver}. In addition to checking the target line is
reached or not,
the Coverage Driver obtains the information about all covered
lines by generated test inputs. 
When an off-target input is generated, the {\it Feedback Generator} 
is scheduled to generate a feedback prompt diagnosing wrong
hypotheses underlying the off-target input generation.

The {\it Feedback Generator} suspects potentially incorrect hypotheses 
in the off-target input generation, and select a set of expressions 
whose runtime evaluations would reveal whether the suspected hypotheses 
actually hold or invalid. 
%
Given selected expressions,
{\it Debugger Driver} formulates a debugging script for an interactive 
debugger (i.e., {\tt gdb}, {\tt lldb}) to extract the runtime values of 
the selected expressions in the off-target test input execution.
Subsequently, the Debugger Driver runs the target program with the
debugging script to record extracted runtime values to a log file.
The Feedback Generator generates a feedback prompt describing which
hypotheses are suspected and what runtime values are obtained from
actual execution.
The Directed Input Generator incorporates given feedback prompt
to refine the context for subsequent test generations (see Section~\ref{sec:feedback}).
These two components are scheduled by the {\it Test Generation 
Controller} to reach the target line efficiently.

\subsection{Motivating Example}
\label{sec:motivating}

One of our case studies with {\tt Poppler} (i.e., Case 2 
in Sec.~\ref{sec:casestudy}) illustrates  
how the proposed technique effectively guides LLM-based 
directed input generation. 
In this study, the proposed technique is given Line 678 in
{\tt addCombining()} of {\tt TestOutputDev.cc} as the directed input generation target.
This test generation process spans five alternating turns of
the Directed Input Generator and Feedback Generator before
reaching the target line.

\begin{itemize}[leftmargin=10pt]

\item {\bf Turn 1. Directed Input Generator:}
    Initially, the proposed technique included the entire code of 
    {\tt addCombining()} in the prompt, along with the code of functions
    in the shortest call path and performed 
    five trials of directed input generation. 
    The program was executed with each inputs and 
    the line coverage was measured. It turned out that 
    none of the inputs pass the sanity check at
    Line 634, causing the function to return immediately.\\

\item {\bf Turn 2. Feedback Generator:}
    Since the target line (i.e., Line 678) remained uncovered, 
    the framework invoked the Feedback Generator, supplying it 
    with each generated test input and the covered lines. 
    The Feedback Generator first requested the LLM to identify 
    critical runtime states that resulted in the unintended path.
    The LLM identified {\tt len}, {\tt wMode}, and 
    {\tt fA->getWMode()} as relevant values to inspect when the program executes Line 634.
    The Debugger Driver ran {\tt gdb} to execute the program with
    each target-missing test input and retrieved these values at
    runtime: for example, {\tt len} was zero. 
    These observations were fed back to the Directed Input Generator.\\

\item {\bf Turn 3. Directed Input Generator:}
    Upon receiving these observed values, the Directed Input Generator 
    refined the prompt and performed additional test generation trials.
    At the second generation, the test input passed the sanity check
    at Line 634, and covered Lines 637–639, 642, and 676. 
    Yet, the refined test input did not satisfy the branch condition
    at Line 676, and the target still remained uncovered.\\
    
\item {\bf Turn 4. Feedback Generator:}    
    For further exploration, the Feedback Generator was invoked
    with the latest generated input. Based on the LLM query
    result, the runtime values of
    {\tt cCurrent}, {\tt u}, {\tt cPrev}, {\tt len}, and 
    {\tt text[len-1]} were extracted at a breakpoint on Line 676.
    This observations revealed that {\tt cPrev} 
    was zero when the test execution did not satisfy the condition at Line 676.\\

\item {\bf Turn 5. Directed Test Generator:}
    The Directed Input Generator incorporated the runtime value information,
    including {\tt cPrev}, into the test generation prompt, and generated
    additional test inputs.
    The first two generated test inputs failed, but the third one 
    satisfied the branch condition at Line 676, and succeeded in covering the target at Line 678.

\end{itemize}

\begin{figure}[t]
\begin{footnotesize}
\begin{verbatim}
633 addCombining (FontInfo *fA, double fsA, Unicode u, ...) {
634  if (len == 0 || wMode != 0 || fA->getWMode() != 0)
635    return;
...
637  Unicode cCurrent = getCombiningChar(u);
638  Unicode cPrev = getCombiningChar(text[len - 1]);
...
642  if(cCurrent != 0 && unicodeTypeAlphaNum(text[len - 1])){
...
676  if(cPrev != 0 && unicodeTypeAlphaNum(u)) {
...
678     maxD = (fA->getAsc()-fA->getDest()) * fsA; /**TARGET**/
\end{verbatim}
\end{footnotesize}
\caption{Target Code Snippet of Motivating Example}
\label{fig:case2}
\end{figure}

\subsection{LLM-based Directed Test Input Generation}
\label{sec:dig}
The Directed Input Generator constructs a prompt and performs 
an LLM inference with it to generate a directed test input. 
Once a new test input is generated, the Directed Input Generator runs 
the target program through the Coverage Driver to check whether the target line is 
reached and to determine which lines are covered by the test input. 
This coverage information is then passed to the Test Generation Controller.

The test generation prompt includes the target function code annotated with 
the target line location (e.g., Line 678 in Figure~\ref{fig:case2}). 
To support reasoning about the target function call context, the prompt also includes 
the entire code of the functions in a shortest call path to the target function. 
The prompt instructs the LLM to utilize this information 
to predict feasible program states at the target function execution. 
To this end, the Directed Input Generator constructs a call graph of the
target program and then identifies a shortest path from the entry function
(e.g., {\tt main()} or {\tt LLVMFuzzerTestOneInput()}) to the target function. 
This analysis explores call paths in a path-insensitive manner,
and resolves indirect calls by assuming that function pointers
sharing the same type and field may alias each other.

A seed input is a previously generated input that covers certain lines in the target function.
When the seed input and its line coverage are available from preceding steps,
the prompt additionally instructs to re-use the seed input elements to preserve
the target function reachability.
Furthermore, when a feedback prompt is provided (see Section~\ref{sec:feedback}, the prompt instructs the LLM to 
generate new inputs by identifying and fixing misaligned elements in the given 
seed input based on the observed runtime value information.

\subsection{Feedback-guided Refinement of Test Generation}
\label{sec:feedback}

The Feedback Generator produces runtime feedback a prompt text 
describing the runtime value examination 
in the test execution failed at covering the target line.
Given off-target test input, the Feedback Generator provides 
the LLM with the input and its line coverage information and then 
ask it to act as the debugging engineer to list prospective runtime values 
for understanding the root causes of the target miss, and 
to suggest modification of the off-target test input to re-orient it 
to the target. To this end, the prompt instructs the LLM to provide 
at most five runtime value queries. A query consists of 
a code location of a {\tt gdb} breakpoint, and an expression for {\tt gdb} 
to retrieve the runtime values at the breakpoint. Inspired by \textsf{AutoSD}~\cite{AutoSD}, we conjecture that the LLM can suspect the conflicts
between the prior hypothesis about the target path condition and 
the runtime observation, and effectively establish new hypothesis to
refine the test input to reach the target location.

To automate runtime value extraction, the Feedback Generator 
synthesizes a {\tt gdb} script in Python, 
and runs the target binary via the Debugger Driver. The Feedback Generator
extracts observed runtime values from the resulting log,
and then incorporates the observations into the feedback supplied
to the next generation iteration. When the breakpoint of a query reaches
multiple times in an execution, the Feedback Generator provides all observed
values as the answer of the query.

The Test Generation Controller dynamically schedules the overall process 
according to a search strategy within the given testing budget. 
In each iteration, the controller dispatches the directed input generation task 
$N$ times, accounting for the non-deterministic nature of LLM-based generation. 
Initially, the Test Generation Controller does not provide any seed input. 
Once a seed input has been selected, however, the Directed Input Generator receives 
it as a reference for generating the next input.

After running tests with the $N$ generated test inputs, the Test Generation Controller
observes their line coverage, and if any generated input covers a previously uncovered line, 
it updates the seed input to the one that covers the largest number of 
previously uncovered lines in the target function (random selection to break ties).

If none of the generated inputs covers a new line, the Test Generation Controller 
performs feedback-guided generation for $N$ times. Specifically, in each iteration,
it dispatches the Feedback Generator to refine the current seed input and then
invokes the Directed Input Generator using the resulting runtime feedback. 
If any of the newly generated $N$ test inputs covers new lines, 
the controller updates the seed to the input that covers previously uncovered lines most. 
Otherwise, the controller restarts the process 
without a seed input to stop exploiting the current search direction.

\vspace{-0.05in}
\subsection{Implementation}

\Tech/ is implemented as 4343 LoC in Python 3,
together with 102 Lines of the prompt templates.
The call path analysis is implemented with Clang LibTooling \cite{Clang:LibTooling} 
which is built on top of the LLVM 18 \cite{LLVM:CGO04}. 
The currently the \Tech/ implementation is developed and tested for
OpenAI APIs. \Tech/ can be configured to use any model provided via
OpenAI APIs such as GPT-5 and GPT-5.4. Note that the design of \Tech/ 
is not specific to certain LLM APIs, but generally applicable. 
\Tech/ uses Docker version 28.1.1 and runs Docker containers on Ubutu 20.04.1. 
The Coverage Driver uses the {\tt llvm-cov} for measuring the line coverage. 

The Debugger Driver generates a debugger script for {\tt gdb} or {\tt lldb}
to conduct executions to extract runtime values. 
A debugger script sets the breakpoint and prints
the indicated expression for each runtime value query. 
The Debugger Driver returns a JSON object containing the list of 
observed values for each query. The list contains
multiple values if the line is reached multiple times. 
The list is empty if the line is never executed, or
the line or expression is invalid. 
\Tech/ uses {\tt gdb} version 8.1.1 or {\tt lldb} version 18,
depending on build configuration.
For each target program, two Docker containers, one for the Coverage Driver and the other for the Debugger Drivers, are built 
and run independently.

%% file: casestudy.tex
\section{Feasibility Studies}
\label{sec:casestudy}

\subsection{Overview}
We conducted case studies to explore the feasibility of 
the proposed framework with real-world C/C++ programs, and 
elaborate how \Tech/ generates
runtime value feedback and utilizes it to refine
directed input generation effectively.

As target projects, we chose {\tt Poppler} commit-1d23101, and {\tt Libsndfile} commit-86c9f92.
which have been used as the reference versions of the target projects in the Magma benchmark~\cite{magma}.
{\tt Libsndfile} is an audio-data reading and writing library supporting various file formats, 
and {\tt Poppler} is a PDF rendering library.
They are open-source C/C++ programs widely studied in software testing research. 

For each project, we performed 
the pilot studies with 120 randomly sampled functions to measure how many lines of them are reached or
unreached by running the naive version of the Directed Test Generator five times,
which provides only the target function code annotated with the target line in the prompt.
Based on the results, we chose {\tt addCombining()} for {\tt Poppler},
and {\tt ogg\_opus\_read\_refill()} for {\tt Libsndfile}.
In {\tt Poppler}, 
{\tt addCombining()} is a Unicode character manipulation
function which adds a combining character 
to a nearby base character when their positions overlap.
In {\tt Libsndfile}, {\tt ogg\_opus\_read\_refill()} 
reads an Ogg file and decodes 
the next Opus packet into an audio buffer.

The two chosen functions showed substantially lower average
target-reaching rates (0.12 {\tt addCombining()},
and 0.10 for {\tt ogg\_opus\_read\_refill()})
than the other functions in their projects.
The two target lines were selected from the lines 
that were never reached in the pilot studies. 
The four case study subjects are as follows:
\begin{itemize}[leftmargin=10pt]
\item[-] Case 1. {\tt Poppler/TextOutputDev.cc:644} 
\item[-] Case 2. {\tt Poppler/TextOutputDev.cc:678} 
\item[-] Case 3. {\tt Libsndfile/ogg\_opus.c:982} 
\item[-] Case 4. {\tt Libsndfile/ogg\_opus.c:1062} 
\end{itemize}

\begin{table}[t]
    \centering
\resizebox{\columnwidth}{!}{%
    \begin{tabular}{ccccccr}
    \hline
      Subject  & Success &  Tests & Turns &  Feedbacks & Values & Token \\
    \hline
     Case 1 &     Yes &  11   &   2   &  5     &  5      &  90765  \\ 
     Case 2 &     Yes &  13   &   2   &  5     &  10     &  93825  \\ 
     Case 3 &     Yes &  7    &   2   &  5     &  3      &  122436      \\ 
     Case 4 &     No  &  15   &   7   &  15    &  13     &  344475      \\      
     \hline         
    \end{tabular}
}
    \caption{Feasibility Study Results}
    \label{tbl:result}
\end{table}

\Tech/ was conducted once for each subject. In the case studies, 
we configured the Test Generation Controller to repeat the same test generation
task five times (i.e., $N=5$).
The call path analysis identified 13 functions in Cases 1 and 2,
and 6 functions in Cases 3 and 4 as the call paths, respectively.
For the four case studies, \Tech/ used OpenAI GPT-5.4 as backbone.
The LLM API costed 3.26 USD in total~\footnote{
GPT-5.4 costs 15 USD per 1M output tokens in July 2026.
}.

Table~\ref{tbl:result} summarizes the directed test input generation
in the four case studies. The ``Tests'' column shows the number of 
generated test inputs. Our framework succeeded in reaching the
target lines for Cases 1--3 within three to five turns (``Turns''), 
where a turn represents one dispatch of either Directed Input Generator or
Feedback Generator involving one LLM invocation.
Before reaching the target, each test generation process involved 
5 to 15 feedback instances (``Feedbacks''), containing 3 to 13
unique runtime values (``Values'').
This result supports the effectiveness of the feedback-guided refinement.

Among the four, three case studies present successful directed
test input generation, while one case study shows a failure (i.e., Case 4).
Table~\ref{tbl:result} shows that, in all success cases,
target-hitting input was not generated at once, but 
feedback generation and refined test generation were performed.
The ``Values'' column shows that, across multiple feedback generations,
the Feedback Generator consistently queries a few variables or
expressions.

\subsection{Case 1. Success After Restart}

\Tech/ successfully generated a target-hitting test input 
after one refinement round (three turns in total) and a subsequent restart. 
The episode proceeded as follows:
\begin{itemize}[leftmargin=10pt]
\item[1)] The Directed Input Generator first generated five inputs. Three inputs reached the target function but not the target line. Since they had identical line coverage, one test input was selected as the seed.
\item[2)] Using the seed and its coverage, the Feedback Generator produced runtime-value feedback. The Directed Input Generator then generated a refined input based on the seed and feedback. 
\item[3)] None of the five refined inputs hit the target line. However, one refined input covers a previously uncovered line and was selected as the new seed.
\item[4)] \Tech/ generated new feedback for this seed and produced five additional refined inputs. None hit the target line or increased coverage, so the Test Generation Controller discarded the seed and restarted the generation process.
\item[5)] After restart, five new inputs were generated without a seed. Like in Step 1, none of these inputs covers the target line, but four of them reached the target function, so one input was selected as a new seed. 
\item[6)] At the first refinement iteration, the Directed Input Generator successfully generated a target-hitting input.
\end{itemize}

In the follow-up analysis, we examined the ten inputs generated at the beginning of the two refinement cycles by observing the LLM trajectory. The explanations on the ten inputs reveal two distinct input-design patterns, and the seeds selected in the two cycles belonged to different patterns.

This case demonstrates that runtime feedback supports exploitation by refining a seed while preserving its underlying design, whereas restarting supports exploration by introducing a substantially different seed. Their combination ultimately led to a on-target test input.

Case 2 is described in detail as the Motivating Example 
(see Section~\ref{sec:motivating}) which follows a process similar to that of Case 1.

\begin{figure}[t]
\begin{footnotesize}
\begin{verbatim}
 976 static int ogg_opus_read_refill (...) {
 977  uint64_t pkt_granulepos ;
 ...
 981  if (odata->pkt_indx == odata->pkt_len) {
 982   nn = ogg_opus_unpack_next_page (...) ; /**TARGET**/
 983   if (nn <= 0)
 984  	 return nn ;
 985  } 
 986  if (odata->pkt_indx == odata->pkt_len)
 987   return 0 ;
....
1032  if (pkt_granulepos <= oopus->pg_pos) {
1033  	oopus->len = nsamp ;
1034  } else {
1035      if (ogg_page_eos (&odata->opage)) ...
\end{verbatim}
\end{footnotesize}
\caption{Target Code Snippet of Case 3}
\end{figure}



\subsection{Case 3. Success After Refinement}
\Tech/ demonstrated that the
Runtime feedback enabled the Directed Input Generator to revise its initial hypothesis 
and produce a on-target test input with Case 3.
The target line 982 is located in {\tt ogg\_opus\_read\_refill()}, 
which processes packets extracted from an Ogg page to refill an audio-decoding buffer. 
The target line loads and unpacks the next page 
when all currently buffered packets have been consumed. 
The refinement process unfolded as follows:

\begin{itemize}[leftmargin=10pt]
\item[1)] The Directed Input Generator initially hypothesized that 
a minimally valid Ogg Opus file containing the required headers 
and an audio packet would cause the target function 
to be invoked with an exhausted packet buffer. 
This assumption was stated explicitly in its explanation:
"when reading begins, packet indices are initially exhausted."
Accordingly, it generated an input for which 
the target branch was expected to be taken during the first relevant invocation.

\item[2)] The generated input was successfully recognized 
and processed as an Ogg Opus file, but it did not cover the target line. 
This result indicated that the failure was not caused by basic input recognition or parsing, 
but by an incorrect assumption about the runtime state at the target condition.

\item[3)] The Test Generation Controller invoked the Feedback
Generator to devise the runtime value queries to identify
the cause of missing the target line in the preceding
test input.
The Feedback Generator produced the runtime value queries
including the following two expressions: the target branch condition,
{\tt odata->pkt\_indx == odata->pkt\_len}, and
the number of packets remaining in the buffer
, {\tt odata->pkt\_len - odata->pkt\_indx}.

\item[4)] The Debugger Driver launched {\tt gdb} with a script to monitor
the two expressions when the execution reaches at Line 981. After the program runs, {\tt gdb} produced  
a debugging log recording the monitoring results. 
The log record shows that 
{\tt odata->pkt\_indx == odata->pkt\_len} was evaluated to
False, 
and {\tt odata->pkt\_len - odata->pkt\_indx} is evaluated to
1,
indicating that that the packet buffer was not exhausted and
that exactly one packet remained when the target function was invoked.
Thus, the function directly processed the buffered packet 
without executing the target line to load another page. 

\item[5)] Given this runtime value feedback, the Directed Input
Generator generated a new test input which successfully reaches
the target line. 
The trajectory of the Directed Input Generator shows that 
the LLM proposed to "provide two separate audio pages after 
the headers" and to "make the first refill consume 
the only buffered packet and then force a second refill attempt when no packets remain." 
This explanation implies that the LLM
effectively identified the misaligned hypothesis, and refined
it considering the observed runtime behaviors.
\end{itemize}


Figure~\ref{fig:case3} summarizes the code difference between
the first (Step 1) and revised (Step 5) test input producing
scripts. The key change in the refined test script is to have
two audio packets in two separate pages, and move 
the end-of-stream indication to the second page.
The revised test script is properly aligned with the explanation in the trajectory of Step 5. 
The produced input will
request the target functions to be invoked twice, such that
the first invocation consumed the already-buffered packet, 
and the subsequent invocation encountered an exhausted packet buffer and executed the target line to load the next page.
This case illustrates that runtime feedback can falsify 
an incorrect hypothesis about the program state 
and guide the Directed Input Generator 
toward a more precise hypothesis.

\lstdefinestyle{diff1}{
    moredelim=**[is][\color{red}]{@}{@}
}
\lstdefinestyle{diff2}{
    moredelim=**[is][\color{ForestGreen}]{*}{*}
}

\begin{figure}[!t]
\centering
\begin{lstlisting}[
    frame=single,
    breaklines=true,
    linewidth=\columnwidth,
    basicstyle=\scriptsize\ttfamily,
    style=diff1,style=diff2]
def ogg_crc(page: bytes} -> int:
  ...

def build_page(packet_data_list, ...):
  ...

@-audio_packet = b'\xF8\xFF\xFE'@
*+audio_packet1 = b'\xF8\xFF\xFE'*
*+audio_packet2 = b'\xFC\xFF\xFE'*

 page0 = build_page([opus_head],0x02,0,serial,0)
 page1 = build_page([opus_tags],0x00,0,serial,1)
@-page2 = build_page([audio_packet],0x04,960,serial,2)@
*+page2 = build_page([audio_packet1],0x00,960,serial,2)*
*+page3 = build_page([audio_packet2],0x04,1920,serial,3)*

@-out = page0 + page1 + page2@
*+out = page0 + page1 + page2 + page3*

with open(output_filename, 'wb') as f:
  f.write(out)
\end{lstlisting}
\caption{Revision of Test Input Producing Script for Case 3}
\label{fig:case3}
\end{figure}

\subsection{Case 4. Diverse Path Exploration via Feedback}
In Case 4, \Tech/ did not succeed to generate a on-target test input 
within the given time bound. Nonetheless, the execution paths explored
by the generated test inputs support that the runtime feedback mechanism 
contributes substantially to maintaining diversity in the search.
The episode proceeded as follows:
\begin{itemize}[leftmargin=10pt]
\item[1)] The Directed Input Generator first generated five inputs, 
none of which reached the target line. In the meantime, the seed is
updated as one of the generated input which covers the greatest 
number of lines in the target function.

\item[2)] The Test Generation Controller run five refinement attempts
but none of the refined test inputs reaches the target line. Through
this round, one of the refined test inputs covers a previously uncovered 
line in the target function, thus replaced the seed input.

\item[3)] This refinement process continued for two more rounds 
(four additional turns), updating the seed after each round.

\item[4)] Since no generated input had increased target-function 
coverage, the Test Generation Controller discarded the seed after
the third round, and restarted the search with a newly generated seed.

\item[5)] \Tech/ repeatedly started over four more times until the time
bound expired, while no on-target test input was produced.
\end{itemize}

We inspected the line and branch coverage achieved by each seed input
to check whether path exploration continued during the test generation
process, or stalled in the middle. We confirmed that every seed input
takes different combinations of branches in the target function. 
We suspect that the runtime feedback effectively guides the Directed 
Input Generator to revise the seed inputs in ways that changed 
the resulting path conditions and progressively explored alternative 
execution paths.

%% file: experiment-secbench.tex
\section{Preliminary Experiments}
\label{sec:experiment}

  \input{table-secbench}

\subsection{Study Design}
\subsubsection{Target Programs}
We assessed whether \Tech/ is
effective at generating crash-reproducing inputs for 59 bug cases from \textsc{Sec-Bench}~\cite{sec-bench, drillagent}, and compared 
the results with \textsc{DrillAgent}. 
\textsc{Sec-Bench} compiles 200 CVEs reported from 29 popular open-source C/C++ projects to benchmark security vulnerability reproduction techniques. 
Each bug case is accompanied with a PoV input given in
the corresponding CVE report. 
\textsc{DrillAgent}, a LLM-based agentic framework 
for reproducing the crash of the given description,
employees 60 bug cases, which cover 18 open-source projects, from \textsc{Sec-Bench} in its experiments with
proprietary LLMs. We employed the same 59 bug cases as the target programs in our experiments. We excluded CVE-2023-27772 of 
{\tt Libiec61850} since its crash-reproduction requires
generating specific command-line arguments, not input data. 

\subsubsection{Experiment Setup}

We re-used the Docker images and PoVs of the 59
bug cases from the \textsc{DrillAgent}'s replication package~\cite{drillagent}.
We revised the build script to make each target binary 
compatible with {\tt llvm-cov} and {\tt lldb}, so that
\Tech/ can interface with it.

In its original evaluation~\cite{drillagent},
the crash reproduction of a bug case is evaluated as
succeeded if a test execution not only fails but also exhibits the same crash type and stack-trace (i.e., a list of function names and line numbers) as the ground-truth. We applied the same evaluation
criteria in this experiment. 
To this end, we captured the exact symptom that each PoV 
exhibits for each bug case before the experiment. 

\subsubsection{\Tech/ Configuration}

We adjusted \Tech/ to the crash-reproducing tasks by including
an explicit statement of the expected crash type, and 
the code snippets of all functions in the ground-truth stack-trace. 
Considering that \textsc{DrillAgent} uses the experiment setup, 
we configured \Tech/ to use OpenAI GPT-5 in all experiments.
%
The rest of the \Tech/ configuration was the same as done in the feasibility studies (Section~\ref{sec:casestudy}).


\subsubsection{Threats to Validity}

This experiment design may not permit a fair comparison of \Tech/ and 
\textsc{DrillAgent} because their implementations run with 
different LLMs under different controls. 
Per target program, \Tech/ calls OpenAI GPT-5 at most 40 times for test input 
and feedback generations (i.e., upto 4 refinement iterations), while \textsc{DrillAgent} uses
Claude Sonnet 4.5 jointly with Haiku 4.5, until it uses up the token cost of USD 1.5. 
Despite these differences, we believe that the validity threat is limited, 
since many LLM benchmark results suggest Claude Sonnet 4.5 and GPT-5 have
comparable capabilities. For example, the SWE-Bench leaderboard~\cite{swebench-web} on 
the {\it Verified} tasks by {\it mini-swe-agent} ranks Claude Sonnet 4.5 and OpenAI GPT-5 (medium) at 13th and 21st places, respectively. 

\Tech/ costs higher (on average USD 3.9) than \textsc{DrillAgent} (at most USD 1.5) when it completes all four iterations. Unlike \textsc{DrillAgent}, the current implementation of \Tech/ does not involve
any particular cost optimization.
We believe that there remain many optimization opportunities for \Tech/
to reduce the token consumption without performance loss.

\subsection{Results}
\label{sec:result}

Table~\ref{tbl:secbench} presents the experiment results of
\Tech/, together with the results of \textsc{DrillAgent} as 
reported in its original evaluation~\cite{drillagent} for reference. 
In the third and fourth columns, `\Vy' indicates that
the corresponding technique succeeded and `\Xn' failed at generating
the crash-reproducing input. In the third column, the number surrounded by 
parenthesis represents the number of refinement iterations conducted 
before generating first crash-reproducing input.  

\Tech/ successfully reproduced crashes in 24 bug cases across 11 projects. 
For seven of these 24 bug cases, \Tech/ required feedback-guided 
refinement at least once (on 2.3 times) before generating the 
crash-reproducing inputs. 
Note that \textsc{DrillAgent} failed in six of these seven bug cases where
\Tech/ took one or more feedback refinement iterations.
We suspect that, in these bug cases, the runtime value feedback 
effectively guided the subsequent test generations toward exploring 
execution paths relevant to the crash reproduction.

Compared with \textsc{DrillAgent}, \Tech/ succeeded on most of the bug cases
(i.e., 13 of the 15 bug cases) where \textsc{DrillAgent} succeeded.
Furthermore, \Tech/ succeeded in 11 additional bug cases across seven projects
where \textsc{DrillAgent} has no success.
The bug cases uniquely reproduced by \textsc{DrillAgent} are only two
from Imagemagick.
Among 15 bug cases where \textsc{DrillAgent} succeeded, 13 were also
reproduced successfully by\Tech/, and 12 of them were reproduced at 
their initial test generation phases, without any feedback generation.

Overall, these results imply that our approach can be considerably 
promising for improving crash-reproduction performance of LLM-based 
directed test input generation, 
although more comprehensive and controlled empirical evaluation should
be supported to 
fairly compare its performance with the existing crash-reproduction techniques.



%% file: table-secbench.tex
\begin{table}
{\footnotesize
    \centering
    \begin{tabular}{c|c|rl|c}
\hline

\hspace{0.5cm}Project\hspace{0.5cm} & 
CVE & 
\multicolumn{2}{c|}{\hspace{0.3cm}\textsc{ReDig}\hspace{0.3cm}$$} &
\textsc{DrillAgent} \\

\cline{1-5}
\cline{1-5}

Exiv2 & CVE-2017-14864 &\hspace{0.4cm} \Xn &  & \Xn   \\
Exiv2 & CVE-2017-17669 & \Vy &\hspace{-0.25cm}(0) \hspace{0.2cm} & \Vy \\
Exiv2 & CVE-2018-17229 & \Vy &\hspace{-0.25cm}(0) & \Vy \\

Faad & CVE-2018-20359 & \Xn  & & \Xn  \\
Faad & CVE-2018-20197 & \Xn  & & \Xn  \\
Faad & CVE-2018-20198 & \Xn  & & \Xn  \\

Gpac & CVE-2021-32437   & \Xn  &    & \Xn  \\
Gpac & CVE-2021-32438   & \Vy &\hspace{-0.25cm}(0) & \Xn  \\
Gpac & CVE-2021-32439   & \Vy &\hspace{-0.25cm}(3) & \Xn  \\
Gpac & CVE-2022-1795    & \Xn  &    & \Xn  \\
Gpac & CVE-2023-0760    & \Vy &\hspace{-0.25cm}(2) & \Vy \\
Gpac & CVE-2023-0819    & \Vy &\hspace{-0.25cm}(0) & \Vy \\
Gpac & CVE-2023-4681    & \Vy &\hspace{-0.25cm}(1) & \Xn  \\
Gpac & CVE-2023-4682    & \Xn  &     & \Xn  \\
Gpac & CVE-2023-4758    & \Vy &\hspace{-0.25cm}(0) & \Vy \\
Gpac & CVE-2023-5998    & \Xn  &     & \Xn  \\
Gpac & CVE-2023-41000   & \Xn  &     & \Xn  \\
Gpac & CVE-2023-46928   & \Vy &\hspace{-0.25cm}(0) & \Vy \\
Gpac & CVE-2023-46930   & \Vy &\hspace{-0.25cm}(2) & \Xn  \\
Gpac & CVE-2023-48011   & \Xn  &     & \Xn  \\

Imagemagick & CVE-2017-11540 & \Xn  &  & \Xn    \\
Imagemagick & CVE-2017-18022 & \Vy &\hspace{-0.25cm}(0) & \Vy \\
Imagemagick & CVE-2019-13299 & \Xn &    & \Vy    \\
Imagemagick & CVE-2019-13309 & \Vy &\hspace{-0.25cm}(0) & \Vy \\
Imagemagick & CVE-2019-13310 & \Vy &\hspace{-0.25cm}(0) & \Vy \\
Imagemagick & CVE-2019-13311 & \Vy &\hspace{-0.25cm}(0) & \Vy \\
Imagemagick & CVE-2019-17541 & \Xn  &  & \Vy   \\
Imagemagick & CVE-2021-3962 & \Vy &\hspace{-0.25cm}(0) & \Vy \\

Mruby & CVE-2018-11743 & \Xn  & & \Xn     \\
Mruby & CVE-2018-12248 & \Xn  & & \Xn    \\
Mruby & CVE-2018-12249 & \Vy &\hspace{-0.25cm}(0) & \Xn  \\
Mruby & CVE-2022-0614  & \Xn  & & \Xn     \\
Mruby & CVE-2022-0717  & \Xn  & & \Xn     \\
Mruby & CVE-2022-1212  & \Xn  & & \Xn     \\
Mruby & CVE-2022-1276  & \Xn  & & \Xn     \\
Mruby & CVE-2022-1427  & \Xn  & & \Xn     \\

Libiec61850 & CVE-2021-45769 & \Xn  &  & \Xn  \\
Libiec61850 & CVE-2023-27772 & - &   & \Xn  \\

Libjpeg-turbo & CVE-2020-13790  & \Xn  & & \Xn  \\

Libmodbus & CVE-2022-0367       & \Vy &\hspace{-0.25cm}(4) & \Xn \\

Libplist & CVE-2017-5545 & \Vy &\hspace{-0.25cm}(0) & \Vy \\

Libredwg & CVE-2020-21814 & \Xn  &  & \Xn  \\
Libredwg & CVE-2020-21816 & \Xn  &  & \Xn  \\
Libredwg & CVE-2020-21827 & \Xn  &  & \Xn  \\
Libredwg & CVE-2020-21834 & \Xn  &  & \Xn   \\
Libredwg & CVE-2021-28236 & \Xn  &  & \Xn  \\
Libredwg & CVE-2021-28237 & \Vy &\hspace{-0.25cm}(2) & \Xn  \\
Libredwg & CVE-2021-42586 & \Xn  &  & \Xn  \\

Libxls & CVE-2023-38855 & \Vy &\hspace{-0.25cm}(0) & \Xn  \\

Matio & CVE-2019-20017  & \Xn  &  & \Xn    \\

Njs & CVE-2019-13617 & \Vy &\hspace{-0.25cm}(2) & \Xn  \\
Njs & CVE-2022-29779 & \Xn  &     & \Xn  \\
Njs & CVE-2022-31306 & \Xn  &     & \Xn  \\
Njs & CVE-2023-27730 & \Xn  &     & \Xn  \\

Openexr & CVE-2020-16587 & \Vy &\hspace{-0.25cm}(0) & \Xn  \\

Openjpeg & CVE-2024-56827 & \Xn  &  & \Xn  \\

Qpdf & CVE-2021-36978     & \Xn  &  & \Xn  \\

Upx & CVE-2023-23457    & \Vy &\hspace{-0.25cm}(0) & \Xn     \\

Yara & CVE-2016-10211   & \Vy &\hspace{-0.25cm}(0) & \Vy     \\
Yara & CVE-2017-5924    & \Xn  &   & \Xn         \\

\cline{1-5}
\cline{1-5}

\multicolumn{2}{r|}{\# Success} & \multicolumn{2}{c|}{24} & 15 \\

\hline

    \end{tabular}
}

\caption{Failure Reproduction Results}

\label{tbl:secbench}
\end{table}

%% file: discussion.tex
\section{Discussion}
\subsection{Related Work}
\textsc{Atlantis}~\cite{atlantis} uses LLMs with program analysis techniques to discover vulnerabilities
and generate PoV inputs.
Its Path-Based PoV Generator iteratively refines LLM-generated inputs using
execution feedback that indicates how far the execution of each input progresses along a path from a 
fuzzing harness to a security-sensitive sink. \textsf{OSS-Fuzz-Gen}~\cite{ossfuzzgen} introduces
context-aware crash validation, which checks whether a crash is feasible
from realistic program entry points by considering its surrounding call and state context.
These works motivate our use of execution feedback and program context
to refine LLM-based directed test input generation.

Recently, more techniques have been proposed to utilize
coverage information as feedback for 
LLM-based test generation.
\textsc{Panta}~\cite{panta} combines static control-flow analysis with dynamic coverage analysis to prioritize under-tested execution paths and asks the LLM to generate tests for them.
\textsc{CoverUp}~\cite{coverup} generates Python regression 
tests targeting uncovered lines and branches. 
When the generated tests produce errors, \textsf{CoverUp} utilizes
the error information to refine the tests such that the intended
test coverage can be attained in the subsequent test generation.

Compared with these existing techniques,
the salient technical aspect of \Tech/ is that
the Feedback Generator of \Tech/ specifically instructs a LLM to 
determine which runtime values to inspect via an interactive
debugger. 
In addition, the Directed Input 
Generator of \Tech/ effectively utilizes both the
runtime value feedback and the previously generated test inputs 
to progressively guide directed input generation toward the target.

\textsc{TestWeaver}~\cite{testweaver} augments the target code with
the runtime values observed in the prior test executions to assist
LLMs to accurately reason about the context of uncovered code regions.
\Tech/ differs from \textsc{TestWeaver} in that it autonomously decides
which runtime values should be observed to determine incorrect 
assumptions, and how the observed runtime values are used in subsequent
test input generations. We believe that the autonomous runtime feedback
generation is crucial for the efficiency and scalability of \Tech/. 
The case studies show that 
\Tech/ is capable of testing C/C++ programs in the 
real world with a large number of functions,
while the experiments of \textsc{TestWeaver} use small-scale
Python programs only.

\subsection{Future Work}
This paper presents our work-in-progress of the \Tech/ framework improving the LLM-based directed input generation via execution-based grounding with runtime value feedback.
Building on the current framework, 
we plan to extend \Tech/ and investigate the following research directions:%

\begin{itemize}[leftmargin=10pt]

\item 
{\bf Exploring alternative feedback-generation strategies.} Currently, the Feedback Generator uses a fixed prompt to instruct the LLM to formulate runtime-value queries for diagnosing misalignment in previously generated test inputs. We plan to investigate alternative prompting strategies inspired by different debugging hypotheses and reasoning patterns. Such diversity may encourage broader exploration of possible failure causes and improve the effectiveness of iterative input refinement.\\

\item
{\bf Reusing previously generated seeds and feedback.} The Directed Test Generator currently uses only the latest seed input as a seed for generating the next test input. Consequently, useful information acquired during earlier test-generation attempts may not be fully utilized. We plan to extend the Test Generation Controller to select and provide multiple relevant seed inputs, thereby enabling more effective knowledge transfer across iterations.\\

\item
{\bf Exploring adaptive search control.} The current Test Generation Controller uses a fixed number of repeated generation attempts to mitigate LLM nondeterminism. However, the appropriate number of repetitions may vary substantially across target programs and target lines, and it is difficult to determine in advance. We therefore plan to develop an adaptive controller that dynamically adjusts the number and type of generation attempts.\\ 

\item
{\bf Comprehensive empirical evaluation.} We plan to evaluate \Tech/ with a diverse set of targets exhibiting different structural and semantic characteristics. The evaluation will examine not only target-reachability or crash-reproduction, but also time and compute efficiencies. Furthermore, we will investigate the contribution of each technical aspect through ablation studies.\\

\item {\bf Applying \Tech/ to different downstream tasks} We plan to extend \Tech/ to
generating test inputs targeting code changes to support patch validation and continuous and regression fuzzing~\cite{bugoss,aflchurn,klooster2023continuous}, and generating test inputs satisfying specific properties to support subsequent
debugging activities~\cite{1dfuzz, bugredux}.\\

\end{itemize}